\documentclass[fleqn,10pt]{wlscirep}

\usepackage{verbatim}
\usepackage{dsfont}
\usepackage{cleveref}
\usepackage{braket}
\usepackage{epstopdf}
\usepackage{cancel}

\newcommand{\up}{\ket{\uparrow_z}}
\newcommand{\down}{\ket{\downarrow_z}}
\newcommand{\tr}{\text{Tr}}

\newcommand{\norm}[1]{\left\lVert #1 \right\rVert}

\newcommand{\beq}{\begin{equation}}
\newcommand{\eeq}{\end{equation}}

\DeclareMathOperator{\Tr}{Tr}

\begin{document}

\title{Analyzing the Quantum Zeno and anti-Zeno effects using optimal projective measurements}

\author[1]{Muhammad Junaid Aftab}
\author[2,*]{Adam Zaman Chaudhry}
\affil[1,2]{School of Science \& Engineering, Lahore University of Management Sciences (LUMS), Opposite Sector U, D.H.A, Lahore 54792, Pakistan}
%\affil[2]{School of Science \& Engineering, Lahore University of Management Sciences (LUMS), Opposite Sector U, D.H.A, Lahore 54792, Pakistan}

\affil[*]{adam.zaman@lums.edu.pk}

\begin{abstract}

Measurements in quantum mechanics can not only effectively freeze the state of the quantum system (the quantum Zeno effect) but also accelerate the time evolution of the system (the quantum anti-Zeno effect). In studies of the quantum Zeno and anti-Zeno effects, a quantum state, usually an excited state of the system, is prepared repeatedly by projecting the quantum state onto the initial state again and again. In this paper, we repeatedly prepare the initial quantum state in a different manner. Instead of only performing projective measurements, we allow unitary operations to be performed, on a very short time-scale, after each measurement. We can then repeatedly prepare the initial quantum state by performing some projective measurement and thereafter, after each measurement, performing a suitable unitary operation to repeatedly end up with the initial state.  Our objective thereafter is to find the projective measurements that need to repeatedly performed such that we obtain the minimum possible effective decay rate of the quantum state. Consequently, the quantum state is maximally protected from its environment. We find that these optimized projective measurements keep on changing depending on the measurement time interval. We explicitly find the projective measurements that need to be performed, as well as the optimized effective decay rate, for a variety of system-environment models such as the population decay model, the pure dephasing model, the spin-boson model, and the large spin-boson model. We find that there can be considerable differences between this optimized effective decay rate and the usual decay rate obtained by repeatedly projecting onto the initial state. In particular, the Zeno and anti-Zeno regimes can be considerably modified.

\end{abstract}

\flushbottom
\maketitle
\thispagestyle{empty}

Rapid repeated measurements can slow down the time evolution of a quantum system, an effect known as the Quantum Zeno effect (QZE) \cite{Sudarshan1977}. It has also been found that if the measurements are not rapid enough, an opposite effect, known as the Quantum anti-Zeno effect (QAZE), can occur - the measurements can actually accelerate quantum transitions \cite{KurizkiNature2000, KoshinoPhysRep2005}. Both the QZE and the QAZE have attracted widespread interest both theoretically and experimentally due to their relevance to the foundations of quantum mechanics as well as possible applications in quantum technologies \cite{FacchiPhysLettA2000,RaizenPRL2001,FacchiPRL2002,BaronePRL2004,ManiscalcoPRL2006,SegalPRA2007,FacchiJPA2008,
WangPRA2008,ManiscalcoPRL2008,ZhengPRL2008,AiPRA2010,FacchiJPA2010,BennettPRB2010,YamamotoPRA2010,ThilagamJMP2010, MilitelloPRA2011,XuPRA2011,ZhangJETP2011,CaoPhysLettA2012,RaimondPRA2012, SmerziPRL2012, WangPRL2013,McCuskerPRL2013,ThilagamJCP2013,ChaudhryPRA2014zeno,StannigelPRL2014, ZhuPRL2014, SchafferNatCommun2014,SignolesNaturePhysics2014, DebierrePRA2015, AlexanderPRA2015, QiuSciRep2015, FanSciRep2015, SlichterNJP2016,Chaudhryscirep2016}. Generally, the analysis of QZE and QAZE for open quantum systems assumes that one continues to repeatedly prepare the initial state of the quantum system by repeatedly projecting the system onto the initial state. However, one can prepare the initial state of the system repeatedly in a different way. Instead of restricting ourselves to performing the same projective measurement again and again, we can instead think of allowing ourselves to perform a unitary operation after each projective measurement. This means that we are then no longer restricted with performing the same projective measurement to repeatedly prepare the initial state. Rather, we can now perform any projective measurement, followed by a suitable unitary operation. We can then use this freedom in the choice of the projective measurement to think about maximizing the survival probability of the system quantum state. Our goal is to find the projective measurements that maximize this survival probability, or, equivalently, minimize the effective decay rate. Thus we arrive at a strategy that allows us to maximally protect the quantum state. We can then study the quantum Zeno and anti-Zeno effects with such optimal projective measurements. 

We start off by considering a single two-level system coupled to an environment. Starting from an arbitrary initial state, our goal is to find the projective measurements followed by the unitary operators that can be used to repeatedly prepare the initial state such that the survival probability of the quantum state is maximized. We find a general condition that allows us to find such projective measurements and unitary operators independent of the particular form of the environment. Namely, we find that the dynamically evolving Bloch vector of the quantum system determines the optimal projective measurement and, in turn, the optimal survival probability. We find that, in general, the optimal projective measurements depend on the time interval between the measurements. This contrasts sharply with the usual scenario where the same projective measurement is performed repeatedly, regardless of the measurement interval. Applications of the condition derived are illustrated by studying the usual decay rate and the optimized decay rate for the population decay model, the pure dephasing model, and the spin-boson model. Since the behaviour of the effective decay rate allows us to identify the Quantum Zeno and anti-Zeno regimes, we can instead use this optimized decay rate to look at the Quantum Zeno and anti-Zeno effects. We then consider a collection of two-level systems interacting with a common environment. In this case, finding the optimal projective measurement is a very difficult problem. However, we can restrict ourselves to measurements that project onto spin coherent states since these are the measurements that can be performed relatively easily experimentally. For this scenario, we again derive the optimal measurements and the optimized decay rate for the pure dephasing case as well as the more general scenario with both dephasing and relaxation present. Once again, we find that there can be signficant differences between the unoptimized decay rate and the optimized decay rate. These differences translate to differences in the QZE and the QAZE.

\section*{Results}

\subsection*{Optimal projective measurements for a single two-level system}
\label{optimalconditions}

Let us begin by considering a single two-level system interacting with an arbitrary environment. The Hamiltonian of the system is $H_S$, the environment Hamiltonian is $H_{B}$, and there is some interaction between the system and the environment that is described by the Hamiltonian $V$. The total system-environment Hamiltonian is thus $H = H_S + H_B + V$. At time $t = 0$, we prepare the system state $\ket{\psi}$. In the usual treatment of the quantum Zeno and anti-Zeno effects, repeated projective measurements described by the projector $\ket{\psi}\bra{\psi}$ are then performed on the system with time interval $\tau$. The survival probability of the system state after one measurement is then $s(\tau) = \tr_{S,B}[(\ket{\psi}\bra{\psi} \otimes \mathds{1}) e^{iH_S\tau}e^{-iH\tau} \rho(0) e^{iH\tau}e^{-iH_S\tau}]$, where $\tr_{S,B}$ denotes taking the trace over the system and the environment, $\rho(0)$ is the initial combined state of the system and the environment, and the evolution of the system state due to the system Hamiltonian itself has been eliminated via a suitable unitary operation just before performing the measurement \cite{MatsuzakiPRB2010,ChaudhryPRA2014zeno,Chaudhryscirep2016}. Assuming that 
the system-environment correlations can be neglected, the survival probability after $N$ measurements can be written as $[s(\tau)]^N = e^{-\Gamma(\tau)N\tau}$, thereby defining the effective decay rate $\Gamma(\tau)$. It should be noted that the behaviour of the effective decay rate $\Gamma(\tau)$ as a function of the measurement interval allows us to identify the Zeno and anti-Zeno regimes. Namely, if $\Gamma(\tau)$ increases as $\tau$ increases, we are in the Zeno regime, while if $\Gamma(\tau)$ decreases if $\tau$ increases, we are in the anti-Zeno regime \cite{KurizkiNature2000, SegalPRA2007, ThilagamJMP2010, ChaudhryPRA2014zeno,Chaudhryscirep2016}. 

We now consider an alternative way of repeatedly preparing the initial state with time interval $\tau$. Once again, we start from the initial system state $\ket{\psi}$. After time $\tau$, we know that the state of the system is given by the density matrix $\rho_S(\tau) = \tr_B[e^{iH_S\tau}e^{-iH\tau} \rho(0) e^{iH\tau}e^{-iH_S\tau}]$, where once again the evolution due to the free system Hamiltonian has been removed. Now, instead of performing the projective measurement $\ket{\psi}\bra{\psi}$, we perform an arbitrary projective measurement given by the projector $\ket{\chi}\bra{\chi}$. The survival probability is then $s(\tau) = \tr_S[(\ket{\chi}\bra{\chi})\rho_S(\tau)]$, and the post-measurement state is $\ket{\chi}$. By performing a unitary operation $U_R$ on the system state on a short timescale, where $U_R\ket{\chi} = \ket{\psi}$, we can again end up with the initial state $\ket{\psi}$ after the measurement. This process can then, as before, repeated again and again to repeatedly prepare the system state $\ket{\psi}$. Once again, if the correlations between the system and the environment can be neglected, we can write the effective decay rate as $\Gamma(\tau) = -\frac{1}{\tau}\ln s(\tau)$. But now, we can, in principle, via a suitable choice of the projector $\ket{\chi}\bra{\chi}$, obtain a larger survival probability (and a correspondingly smaller decay rate) than what was obtained with repeatedly using projective measurements given by the projector $\ket{\psi}\bra{\psi}$. The question, then, is what is this projector $\ket{\chi}\bra{\chi}$ that should be chosen to maximize the survival probability?

For an arbitrary quantum system, it is difficult to give a general condition or formalism that will predict this optimal projective measurement. However, most studies of the effect of repeated quantum measurements on quantum systems have been performed by considering the quantum system to be a single two-level system \cite{KoshinoPhysRep2005}. Let us now show that if the quantum system is a two-level system, then it is straightforward to derive a general method for calculating the optimal projective measurements that need to be performed as well as an expression for the optimized decay rate. We start from the observation that the system density matrix at time $\tau$, just before the measurement, can be written as 
\begin{equation} \label{do}
\rho_{S}(\tau) = \frac{1}{2} \Big (\mathds{1} + n_{x}(\tau)\sigma_{x} + n_{y}(\tau)\sigma_{y} + n_{z}(\tau)\sigma_{z} \Big ) = \frac{1}{2} \Big ( \mathds{1} + \mathbf{n}(\tau) \cdot \mathbf{\sigma} \Big ),
\end{equation}
where $\mathbf{n}(\tau)$ is the Bloch vector of the system state.
We are interested in maximizing the survival probability $s(\tau) = \tr_S[(\ket{\chi}\bra{\chi})\rho_S(\tau)]$. It is clear that we can also write 
\begin{equation} \label{dop}
\ket {\chi} \bra {\chi} = \frac{1}{2} \Big ( \mathds{1} + n'_{x}\sigma_{x} + n'_{y}\sigma_{y} + n'_{z}\sigma_{z} \Big ) = \frac{1}{2} \Big ( \mathds{1} + \mathbf{n'} \cdot \mathbf{\sigma} \Big ),
\end{equation}
where $\mathbf{n'}$ is a unit vector corresponding to the Bloch vector for the projector $\ket{\chi}\bra{\chi}$. Using Eqs.~\eqref{do} and \eqref{dop}, we find that the survival probability is 
\begin{equation}
s(\tau) = \frac{1}{2}\left(1 + \mathbf{n}(\tau) \cdot \mathbf{n'} \right). 
\end{equation}
It should then be obvious how to find the optimal projective measurement $\ket{\chi}\bra{\chi}$ that needs to be performed. The maximum survival probability is obtained if $\mathbf{n'}$ is parallel to $\mathbf{n}(\tau)$. If we know $\rho_S(\tau)$, we can find out $\mathbf{n}(\tau)$. Consequently, $\mathbf{n'}$ is simply the unit vector parallel to $\mathbf{n}(\tau)$. Once we know $\mathbf{n'}$, we know the projective measurement $\ket{\chi}\bra{\chi}$ that needs to be performed. The corresponding optimal survival probability is given by 
\begin{equation}
\label{optimizedprobability}
s^{*}(\tau) = \frac{1}{2}\left(1 + \norm{\mathbf{n}(\tau)}\right). 
\end{equation}
Now, if we ignore the correlations between the system and environment, which is valid for weak system-environment coupling, we can again derive the effective decay rate of the quantum state to be $\Gamma(\tau) = -\frac{1}{\tau}\ln s^{*}(\tau)$. We now investigate the optimal effective decay rate for a variety of system-environment models.

\subsection*{The population decay model}

\begin{figure}
{\includegraphics[scale = 0.55]{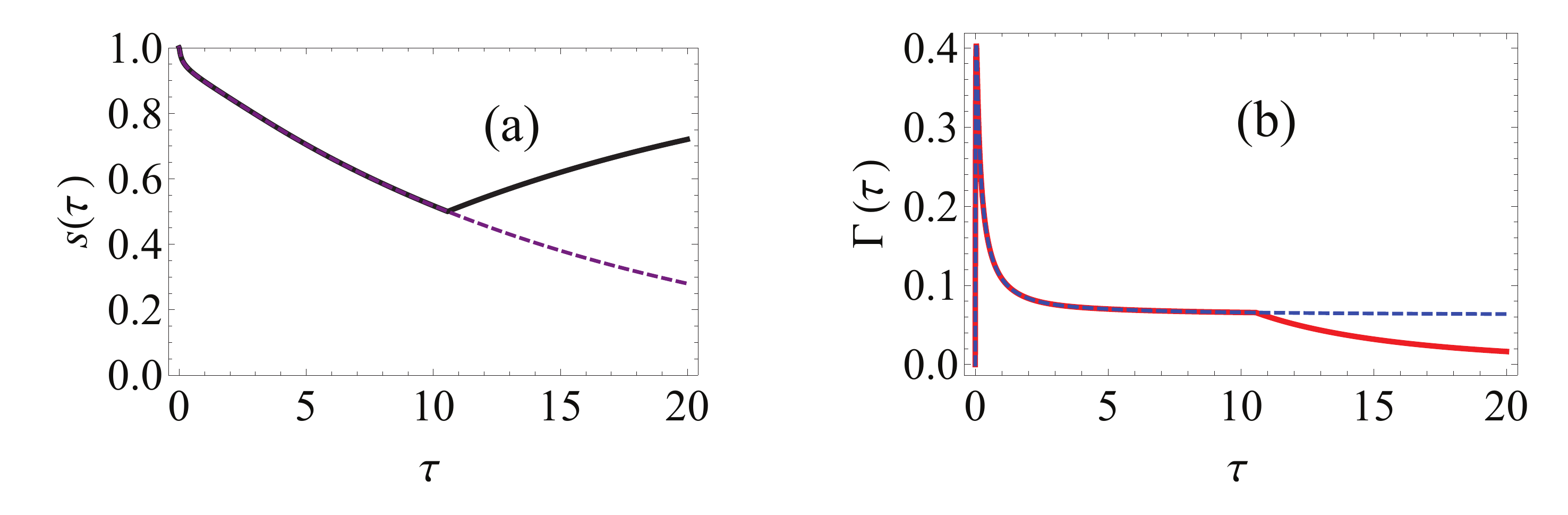}}\caption{\textbf{Behaviour of both the survival probability and the effective decay rate for the population decay model.} \textbf{(a)} $s(\tau)$ versus $\tau$. The purple dashed curve shows the survival probability if the excited state is repeatedly measured; the black curve shows the survival probability if the optimal projective measurement is repeatedly made. \textbf{(b)} $\Gamma(\tau)$ versus $\tau$. The blue dashed curve shows the decay rate if the excited state is repeatedly measured; the solid red curve shows the decay rate if the optimal projective measurement is repeatedly made. We have used $G = 0.01, \omega_c = 50$ and $\varepsilon = 1$. In this case, $\tau^* \approx 10.6$.}
\label{PopDecayMerged2}
\end{figure}

To begin, we consider the paradigmatic population decay model. The system-environment Hamiltonian is (we use $\hbar = 1$ throughout)
\begin{equation}
H = \frac{\varepsilon}{2}\sigma_{z} + \sum_{k} \omega_{k} b_{k}^{\dagger} b_{k} + \sum_{k} (g_{k}^{*}b_{k}\sigma^{+} + g_{k}b_{k}^{\dagger}\sigma^{-}), 
\end{equation}
where $\varepsilon$ is the energy difference between the two levels, $\sigma_{z}$
is the standard Pauli matrix, $\sigma^{+}$ and $\sigma^{-}$ are the raising
and lowering operators, and $b_k$ and $b_k^{\dagger}$ are the annihilation and creation operators for mode $k$ of the environment. It should be noted that here we have made the rotating-wave approximation. This system-environment Hamiltonian is widely used to study, for instance, spontaneous emission \cite{Scullybook}. We consider the very low temperature regime. We initially prepare the system-environment state $\ket{\uparrow_z,0}$ that describes the two-level system to be in the excited state and the environment oscillators to be in their ground state. Ordinarily, in the studies of the QZE and the QAZE, the system is repeatedly projected onto the excited state with time interval $\tau$. As discussed before, we,  on the other hand, allow the system to be projected onto some other state such that the effective decay rate is minimized. To find this optimal projective measurement, we need to understand how the Bloch vector of the system evolves in time. Due to the structure of the system-environment Hamiltonian, the system-environment state at a later time $\tau$ can be written as $
 | \psi(t) \rangle = f(t) \ket{\uparrow_z, 0} + \sum_{k} f_k(t) \ket{\downarrow_z, 0}$,
where $\ket{\downarrow_z, k}$ means that the two-level system is in the ground state and that
mode $k$ of the environment has been excited. It then follows that the density matrix of the system at time $\tau$ is 
\begin{align}\label{reduceddebsitymatrix}
\rho_{S}(\tau)& = 
\begin{bmatrix}
|f(t)|^2 & 0 \\
0 & \displaystyle \sum_{k} |f_{k}(t) |^2
\end{bmatrix}
\end{align}

We consequently find that the components of the Bloch vector of the system are $n_{x}(t) = n_{y} (t) = 0$, while $n_{z}(t) = 1 - 2 \displaystyle \sum_{k} |f_{k}(t) |^2$.
\begin{comment}
\begin{align}\label{nzpop}
n_{z}(\tau) & = \Tr (\sigma_{z} \rho(\tau)), \nonumber \\
& = \Tr \begin{bmatrix}
1 & 0 \\
0 & -1
\end{bmatrix} 
\begin{bmatrix}
|f(\tau)|^2 & 0 \\
0 & \displaystyle \sum_{k} |f_{k}(\tau) |^2
\end{bmatrix} \nonumber, \\
& = \Tr \begin{bmatrix}
|f(\tau)|^2 & 0 \\
0 & -  \displaystyle \sum_{k} |f_{k}(\tau) |^2
\end{bmatrix} \nonumber, \\ 
& = |f(\tau)|^2 -  \displaystyle \sum_{k} |f_{k}(\tau) |^2, \nonumber \\
& = 1 - 2 \displaystyle \sum_{k} |f_{k}(\tau) |^2,
\end{align}
\end{comment}
Thus, we have a nice interpretation for the dynamics of the system. Initially, the system is in the excited state. As time goes on, the coherences remain zero, and the probability that the system makes a transition to the ground state increases. In other words, initially the Bloch vector of the system is a unit vector with $n_z(0) = 1$. The Bloch vector then decreases in magnitude (while keeping the $x$ and $y$ components zero) until the size of the Bloch vector becomes zero. The Bloch vector thereafter flips direction and increases in length until effectively $n_z(t) = -1$. Since the direction of the Bloch vector corresponding to the optimal measurement is parallel to the Bloch vector of the system, we find that if the measurement interval is short enough such that $n_z(\tau) > 0$, then we should keep on applying the projector $\ket{\uparrow_z}\bra{\uparrow_z}$. On the other hand, if the measurement interval is large enough so that $n_z(\tau) < 0$, then we should rather apply the projector $\ket{\downarrow_z}\bra{\downarrow_z}$, and then, just after the measurement, apply a $\pi$ pulse so as to end up with the system state $\ket{\uparrow_z}$ again. In other words, the time $t = \tau^*$ at which the Bloch vector flips direction is of critical importance to us and needs to be found out in order to optimize the effective decay rate. To find this time, we assume that the system and the environment are weakly coupled and thus we can use a master equation to analyze the dynamics of the system. Since we numerically solve this master equation, we might as well put back in the non-rotating wave approximation terms so that the system-environment interaction that we consider in solving the master equation is $\sum_{k} \sigma_{x} (g_{k}^{*}b_{k} + g_{k}b_{k}^{\dagger})$. The master equation that we use can be written as   
\begin{equation} \label{masterequation}
\frac{d \rho_S (t)}{dt} = i[\rho_S(t), H_{S}] + \int_{0}^{t} ds \Big \{ [\bar{F}(t,s)\rho_S(t), F ]C_{t s} + \; \text{h.c.} \Big \},
\end{equation}
where the system Hamiltonian is $H_S = \frac{\varepsilon}{2}\sigma_z$, the system-environment interaction Hamiltonian has been written as $F \otimes B$ with $F = \sigma_x$ and $B = \sum_{k} (g_{k}^{*}b_{k} + g_{k}b_{k}^{\dagger})$, $\bar{F}(t,s) = e^{iH_S(t -s)}Fe^{-iH_S(t -s)}$, and h.c. denotes the hermitian conjugate. Here the environment correlation function $C_{ts}$ is defined as $C_{ts} = \tr_B [\widetilde{B}(t)\widetilde{B}(s) \rho_B]$ where $\widetilde{B}(t) = e^{iH_B t} B e^{-iH_B t}$ and $\rho_B = e^{-\beta H_B}/Z_B$ with $Z_B$ the partition function. To find the environment correlation function, we introduce the spectral density function $J(\omega) = G \omega^s \omega_{c}^{1-s} e^{- \omega/ \omega_c}$, where $G$ parametrizes the system-environment coupling strength, $s$ characterizes the Ohmicity of the
environment, and $\omega_c$ is the cutoff frequency. We can then numerically solve this differential equation to find the system density matrix at any time $t$ and consequently the Bloch vector of the system. We consequently know the optimal projective measurement that needs to performed.

\begin{figure}
{\includegraphics[scale = 0.5]{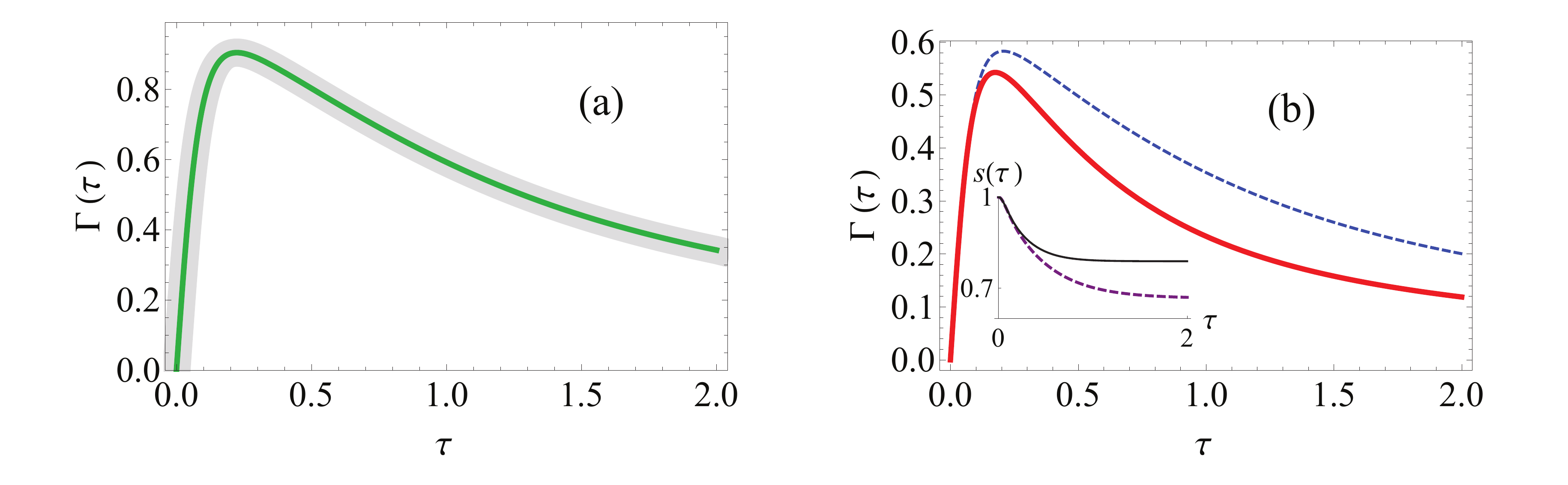}}
\caption{\label{Dephasing1}\textbf{Graphs of the effective decay rate under making optimal projective measurements in the pure dephasing model.} \textbf{(a)} $\Gamma(\tau)$ versus $\tau$ for the initial state specified by the Bloch vector $(1, 0, 0)$. The thickened light gray curve curve shows the decay rate if the initial state is repeatedly measured; the green curve shows the decay rate if the optimal projective measurement is repeatedly made. It is clear from the figure that the two curves identically overlap. \textbf{(b)} $\Gamma(\tau)$ versus $\tau$ for the initial state specified by the Bloch vector $(1/\sqrt{3}, 1/\sqrt{3}, 1/\sqrt{3} )$. The blue dashed curve shows the decay rate if the initial state is repeatedly measured; the solid red curve shows the decay rate if the optimal projective measurement is repeatedly made. We have used $G = 0.1, \omega_c = 10, \beta = 0.5$. For $\tau = 1$ and $N = 3$, the difference in the survival probabilities is already $0.15$.}
\end{figure}
We now present a computational example (see the previous page). We plot the single measurement survival probability [see Fig.~\ref{PopDecayMerged2}(a)] and the effective decay rate [see Fig.~\ref{PopDecayMerged2}(b)] as a function of the measurement interval $\tau$. The dotted lines illustrate what happens if we keep on projecting the system state onto $\ket{\uparrow_z}$. For a small measurement interval, the optimal measurement is $\ket{\uparrow_z}\bra{\uparrow_z}$ since the system Bloch vector has only a positive $z$-component. On the other hand, if we realize that when the measurement interval is long enough such that, between each measurement, the Bloch vector flips direction, then to maximize the survival probability, we should project instead onto the state $\ket{\downarrow_z}$ and then apply $\pi$ pulse. Doing so, we can obtain a higher survival probability or, equivalently, a lower effective decay rate. This is precisely what we observe from the figure. For this population decay case, we find that if the measurement interval is larger than $\tau^* \approx 10.6$, then we are better off by performing the measurement $\ket{\downarrow_z}\bra{\downarrow_z}$. There is also a small change in the anti-Zeno behaviour. For our modified strategy of repeatedly preparing the quantum state, we find that beyond measurement interval $\tau = \tau^*$, there is a sharper signature of anti-Zeno behaviour as compared to the usual strategy of repeatedly measuring the excited state of the system.

\subsection*{Pure dephasing model} \label{dephasing}
\begin{figure}
{\includegraphics[scale = 0.6]{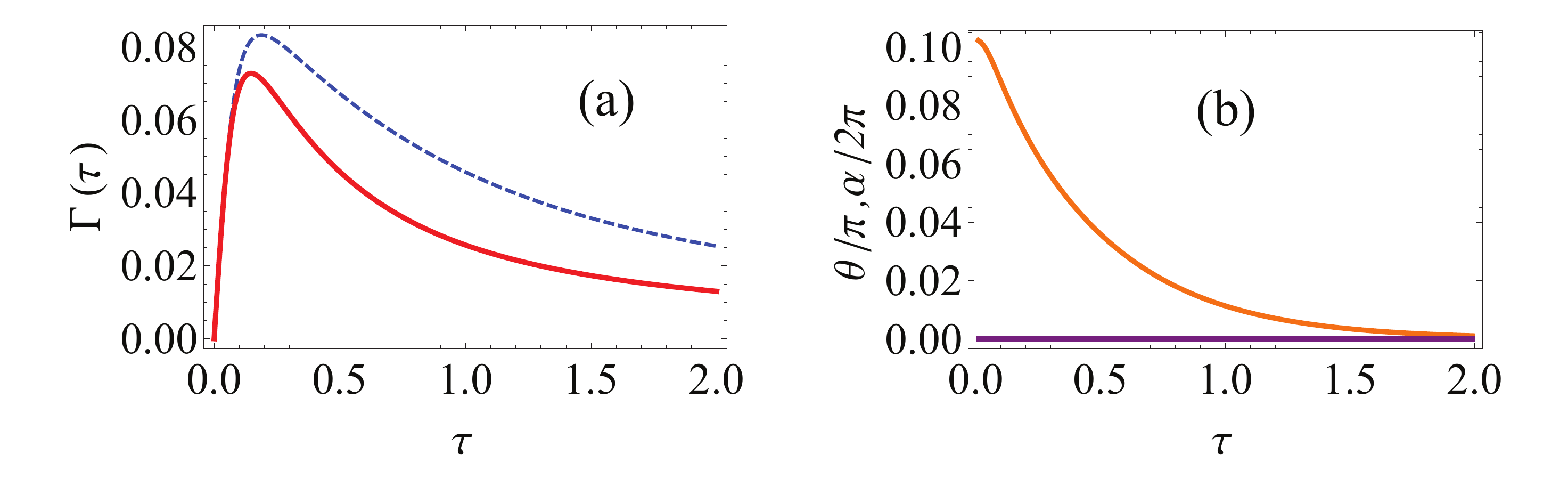}}
\caption{\label{Dephasing2}\textbf{Graphs of the effective decay rate and the optimal spherical angles under making optimal projective measurements in the pure dephasing model.} \textbf{(a)} $\Gamma(\tau)$ versus $\tau$ for the initial state specified by the Bloch vector $(1/\sqrt{10}, 0, \sqrt{9/10} )$. The same parameters as used in Fig.~\ref{Dephasing1} have been used for this case as well. \textbf{(b)} Graphs of the optimal spherical angles that maximize the survival
probability. $\theta$ is the polar angle and $\alpha$ is the azimuthal angle, which remains $0$ at all times.}
\end{figure}
We now analyze our strategy for the case of the pure dephasing model \cite{BPbook}. The system-environment Hamiltonian is now given by
\begin{equation}
H = \frac{\varepsilon}{2}\sigma_{z} + \sum_{k} \omega_{k} b_{k}^{\dagger} b_{k} + \sigma_{z} \sum_{k} (g_{k}^{*}b_{k} + g_{k}b_{k}^{\dagger}).
\end{equation}
The difference now compared to the previous population decay model is that the system-environment coupling term now contains $\sigma_z$ instead of $\sigma_x$. This difference implies that the diagonal entries of the system density matrix $\rho_S(t)$ (in the $\ket{\uparrow_z}, \bra{\uparrow_z}$ basis) cannot change - only dephasing can possibly take place, which is the reason that this model is known as the pure dephasing model\cite{ChaudhryPRA2014zeno}. Furthermore, this pure dephasing model is exactly solvable. The off-diagonal elements of the density matrix undergo both unitary time evolution due to the system Hamiltonian and non-unitary time evolution due to the coupling with the environment. Assuming the initial state of the total system is the standard product state $\rho_S(0) \otimes \rho_B/Z_B$, the off-diagonals of the density matrix $[\rho_S(t)]_{mn}$, once the evolution due to the system Hamiltonian itself is removed, are given by $[\rho_S(t)]_{mn} =  [\rho_{S}(0)]_{mn} e^{- \gamma(\tau)}$ where $\gamma(t) = \sum_k 4|g_k|^2 \frac{(1 - \cos (\omega_k t))}{\omega_k^2} \coth \left( \frac{\beta \omega_k}{2} \right)$\cite{ChaudhryPRA2014zeno}. Writing an arbitrary initial state of the system as $\ket \psi = \cos \Big (\frac{\theta}{2} \Big ) \ket{\uparrow_z} + \sin \Big (\frac{\theta}{2} \Big ) e^{i \phi} \ket{\downarrow_z}$, it is straightforward to find that 
\begin{align} \label{DephasingBlochVector}
n_x(t) = e^{- \gamma(t)} n_{x}(0) , \; n_y(t) = e^{- \gamma(t)} n_{y}(0), \; n_z(t) & = n_{z}(0).
\end{align}
The optimal survival probability obtained using optimized measurements is then 
\begin{equation}\label{probnew}
s^{*}(\tau) = \frac{1}{2} \Big (1 + \sqrt{n_z(0)^2 + (e^{- \gamma(\tau)})^2(n_x(0)^2 + n_z(0)^2)} \; \Big ),
\end{equation}
where Eq.~\eqref{optimizedprobability} has been used. On the other hand, if we keep on preparing the initial state $\ket{\psi}$ by using the projective measurements $\ket{\psi}\bra{\psi}$, we find that 
\begin{equation} \label{probold}
s(\tau) = \frac{1}{2} \Big (1 + n_z(0)^2 + e^{- \gamma(\tau)} \big ( n_x(0)^2 + n_z(0)^2 \big ) \; \Big ).
\end{equation}

We now analyze Eqs.~\eqref{probnew} and \eqref{probold} to find conditions under which we can lower the effective decay rate by using optimized projective measurements. It is clear that if the initial state, in the Bloch sphere picture, lies in the equatorial plane, then $n_z(0) = 0$ while $n_x(0)^2 + n_y(0)^2 = 1$. In this case, Eqs.~\eqref{probnew} and \eqref{probold} give the same survival probability. Thus, in this case, there is no advantage of using our strategy of optimized measurements as compared with the usual strategy. The reasoning is clear. In the Bloch sphere picture,  the magnitude of the time evolved Bloch vector of the density matrix reduces such that the time evolved Bloch vector is always parallel to the Bloch vector of the initial pure system state. As argued before, the optimal projector to measure at time $\tau$, $\ket \chi \bra \chi$, must be parallel to the Bloch vector of the density matrix at time $\tau$. Hence in this case, the optimal projector to measure is $\ket \psi \bra \psi$, corresponding to the initial state. The computational example shown in Fig.~\ref{Dephasing1}(a) illustrates our predictions. 

On the other hand, if we make some other choice of the initial state that we repeatedly prepare, we can easily see that our optimized strategy can give us an advantage. We simply look for the case where the evolved Bloch vector (after removal of the evolution due to the system Hamiltonian) no longer remains parallel with the initial Bloch vector. Upon inspecting Eq.~\eqref{DephasingBlochVector}, we find that our optimized strategy can be advantageous if $n_z(0) \neq 0$ (excluding, of course, the cases $n_z(0) = \pm 1$). In other words, if the Bloch vector of the initial state does not lie in the equatorial plane, then the Bloch vector of this state at some later time will not remain parallel to the initial Bloch vector. In this case then, our optimal measurement scheme will give a higher survival probability as compared to repeatedly measuring the same initial state. This is illustrated in Fig.~\ref{Dephasing1}(b) where we show the effective decay rate and the survival probability after a single measurement for the initial state specified by the Bloch vector $(1/\sqrt{3}, 1/\sqrt{3}, 1/\sqrt{3} )$. After the time at which the transition between the Zeno and the anti-Zeno regimes occurs, we clearly observe that the decay rate is lower when one makes the optimal projective measurements. Although this difference on first sight may not appear as very significant, if we perform a relatively large number of repeated measurements, the difference is very significant. For example, even for three measurements with measurement interval $\tau = 1$, we find that the quantum state has $0.15$ greater survival probability with the optimized measurements as compared with the usual unoptimized strategy of repeatedly preparing the quantum state.

Another computational example has been provided in Fig.~\ref{Dephasing2} where the initial state is now given by the Bloch vector $(1/\sqrt{10}, 0, \sqrt{9/10} )$. In Fig.~\ref{Dephasing2}(a) we have again illustrated that our optimized strategy of repeatedly preparing the quantum state is better at protecting the quantum compared to the usual strategy. In Fig.~\ref{Dephasing2}(b) we have shown how the optimal projective measurement that needs to be performed changes with the measurement interval $\tau$. In order to do so, we have parametrized the Bloch vector corresponding to $\ket{\chi}\bra{\chi}$ using the usual spherical polar angles $\theta$ and $\alpha$. Note that the value of the azimuthal angle $\alpha$ is expected to remain constant since we have 
$\alpha(\tau) = \arctan ( n_{y}(\tau)/n_{x}(\tau) ) = \alpha(0)$.
On the other hand, the optimal value of the polar angle $\theta$ changes with the measurement interval. This is also expected since as the system dephases, $e^{- \gamma(\tau)} \rightarrow 0$, ensuring that $n_{x}(\tau), n_{y}(\tau) \rightarrow 0$. Thus, for long measurement intervals, the system Bloch vector becomes effectively parallel to the $z-\text{axis}$. It follows that $\theta \rightarrow 0$ for long measurement intervals. These predictions are borne out by the behaviour of $\theta$ and $\alpha$ in Fig.~\ref{Dephasing2}(b). 

\subsection*{The Spin-Boson Model}
We now consider the more general system-environment model given by the Hamiltonian
\begin{figure}
{\includegraphics[scale = 0.55]{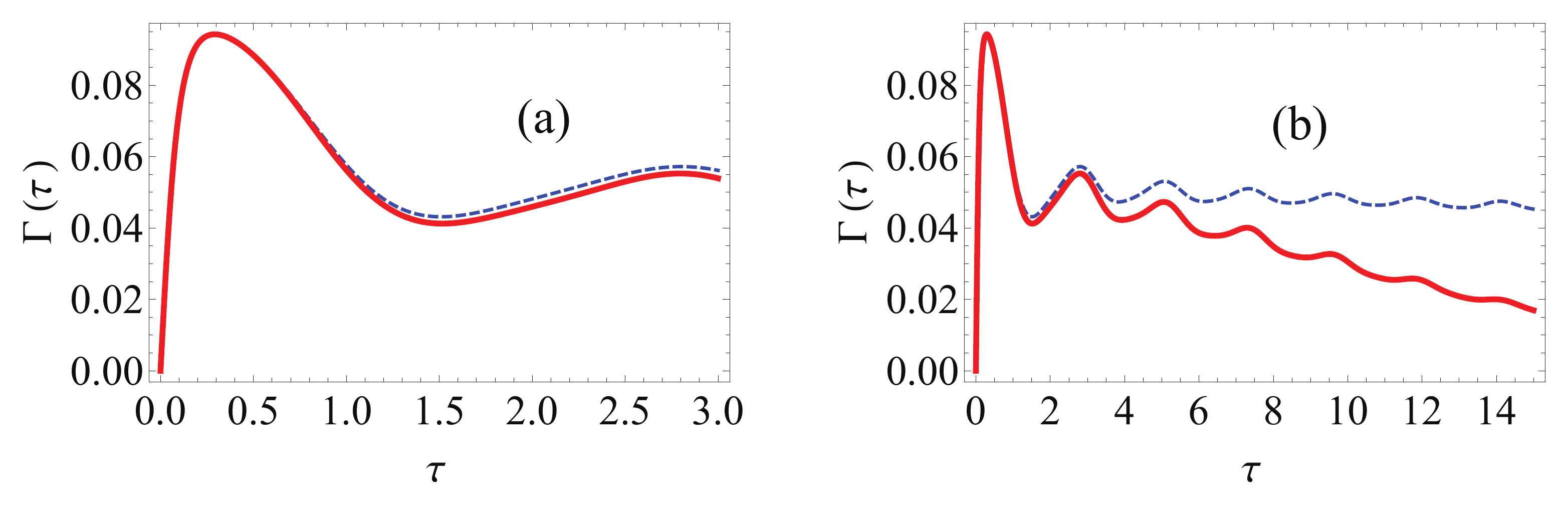}}\caption{\textbf{Graphs of the effective decay rate under making optimal projective measurements in the spin-boson model.} \textbf{(a)} $\Gamma(\tau)$ versus $\tau$ (low temperature) for the state specified by $\theta = \pi/2$ and $\alpha = 0$. The blue dashed curve shows the decay rate in the spin boson model ($\Delta = 2, \varepsilon = 2$) if the initial state is repeatedly measured, and the solid red curve shows the effective decay rate with the optimal measurements. We have used $G = 0.01, \omega_c = 10$ and $s = 1$. \textbf{(b)} is the same as \textbf{(a)} except for the domain of the graph.}
\label{SB1} 
\end{figure}
\begin{equation}\label{spinboson}
H = \frac{\varepsilon}{2} \sigma_z + \frac{\Delta}{2} \sigma_x + \sum_{k} \omega_{k} b_{k}^{\dagger} b_{k} + \sigma_{z} \sum_{k} (g_{k}^{*}b_{k} + g_{k}b_{k}^{\dagger}),
\end{equation}
where $\Delta$ can be understood as the tunneling amplitude for the system, and the rest of the parameters are defined as before.
This is the well-known spin-boson model\cite{LeggettRMP1987,Weissbook,BPbook},  which can be considered as an extension of
the previous two cases in that we can now generally have
both population decay and dephasing taking place. Experimentally, such a model can be realized, for instance, using superconducting qubits \cite{ClarkeNature2008, YouNature2011,SlichterNJP2016} and the properties of the environment can be appropriately tuned as well \cite{HurPRB2012}. Once again, assuming that the system and the environment are interacting weakly with each other, we can use the master equation that we have used before [see Eq.~\eqref{masterequation}] to find the system density matrix as a function of time. We now have $H_S = \frac{\varepsilon}{2}\sigma_z + \frac{\Delta}{2}\sigma_x$ and $F = \sigma_z$. It should be remembered that once we find the density matrix just before the measurement $\rho_S(\tau)$, we remove the evolution due to system Hamiltonian via $\rho_S(\tau) \rightarrow e^{iH_S \tau} \rho_S(\tau)e^{-iH_S\tau}$.  

Let us first choose as the initial state $n_x(0) = 1$ (or, in the words, the state that is paramterized by $\theta = \pi/2$ and $\alpha = 0$ on the Bloch sphere). In Fig.~\ref{SB1}, we plot the behaviour of the effective decay rate as a function of the measurement interval using both our optimized strategy (the solid red lines) and the unoptimized usual strategy (the dotted, blue curves). It is clear from Fig.~\ref{SB1}(a) that for relatively short measurement intervals, there is little to be gained by using the optimal strategy. As we have seen before in the pure dephasing case, for the state in the equatorial plane of the Bloch sphere, there is no advantage to be gained by following the optimized strategy. On the other hand, for longer time intervals $\tau$, population decay can be considered to become more significant. This then means we that see significant difference at long measurement intervals if we use the optimized strategy. This is precisely what we observe in Fig.~\ref{SB1}(b).

\begin{figure}
{\includegraphics[scale = 0.4]{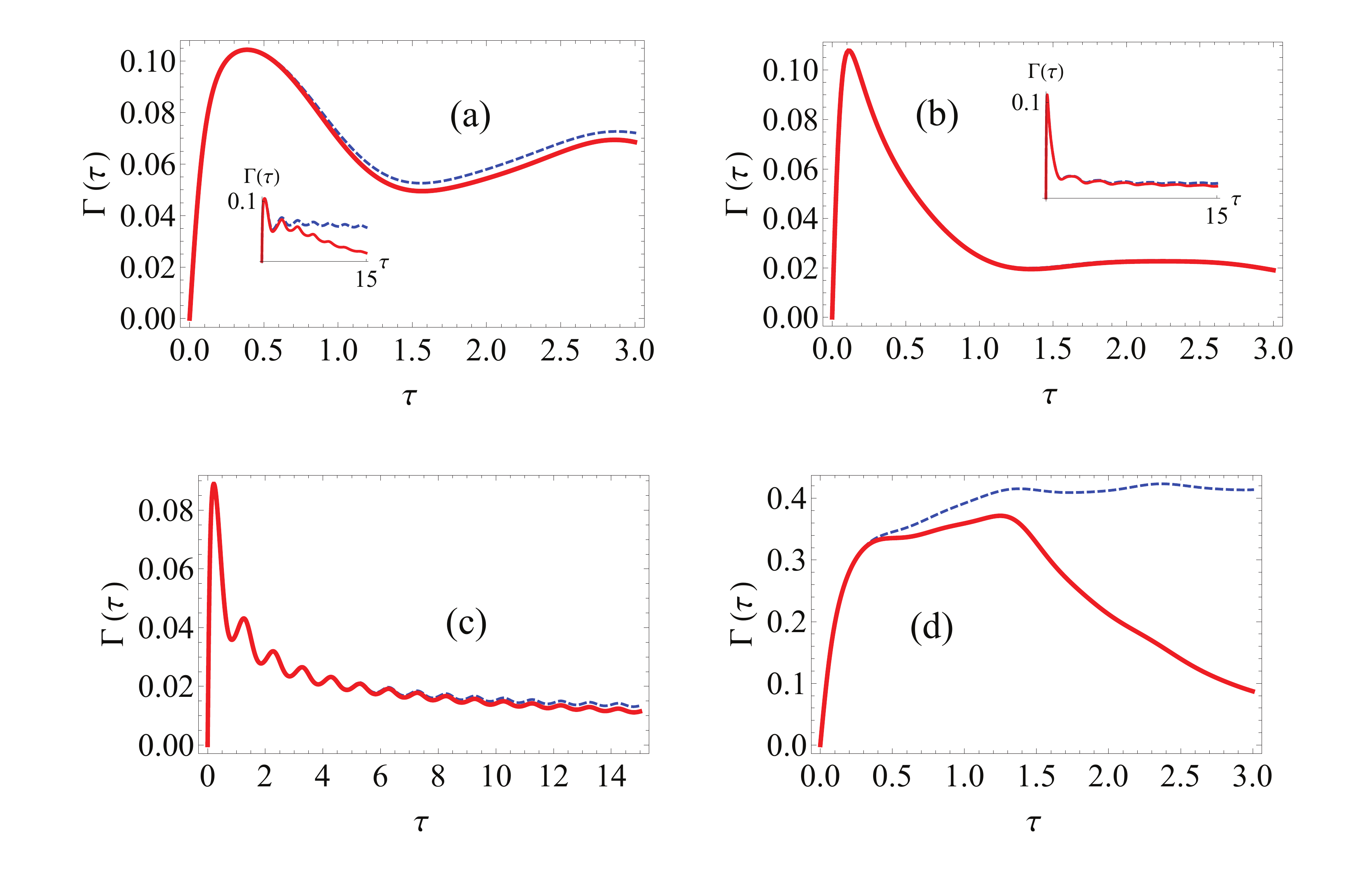}}\caption{\label{SB2} \textbf{Graphs of the effective decay rate under making optimal projective measurements in the spin-boson model.} \textbf{(a)} $\Gamma(\tau)$ versus $\tau$ (low temperature) for the state specified by $\theta = \pi/2$ and $\phi = 0$. We have used the same parameters as in \ref{SB1} except that we have now modeled a sub-Ohmic environment with $s = 0.8$. \textbf{(b)} same as \textbf{(a)}, except that we have now modeled a super-Ohmic environment with $s = 2.0$. \textbf{(c)} We have used  $G = 0.025, \omega_c = 10$ and $s = 1$. For this case ($\varepsilon \gg \Delta$), we have used $\varepsilon = 6, \Delta = 2$. \textbf{(d)}  same as \textbf{(c)}, except that for this case ($\Delta \gg \varepsilon $), we have used $\varepsilon = 2, \Delta = 6$.}
\end{figure}

For completeness, let us also investigate how the the effective decay rate depends on the functional form of the spectral density. In Fig.~\ref{SB2}(a) and Fig.~\ref{SB2}(b), we investigate what happens in the case of a sub-Ohmic and a super-Ohmic environment respectively.  The case of a sub-Ohmic environment with $s = 0.8$ is similar to the case with $s = 1$ (Ohmic environment) - once again, the optimal projective measurements decrease the decay rate substantially only at long periods of time. For the case of a super-Ohmic environment with $s=2$ [see Fig.~\ref{SB2}(b)], we find that the optimal projective measurements do not substantially lower the decay rate, even for long times. Thus, it is clear that the Ohmicity of the environment plays an important role in determining the usefulness of using the optimal projective measurements. 

Let us now revert to the Ohmic environment case to present more computational examples. First, if $\varepsilon \gg \Delta$, then the effect of dephasing is much more dominant than the effect of population decay. Results for this case are illustrated in Fig.~\ref{SB2}(c). We see that there is negligible difference upon using the optimal measurements. This agrees with what we found when we analyzed the pure dephasing model. We also analyze the opposite case where the effect of population decay is more dominant than the effect of dephasing. This is done by setting $\Delta \gg \varepsilon$. We consider higher temperature in this case \underline{what value???}. We now observe differences between the unoptimized and optimized decay rates for relatively short times [see Fig.~\ref{SB2}(d)], and the difference becomes even bigger at longer times. In fact, while we observe predominantly only the Zeno effect with the unoptimized measurements, we observe very distinctly both the Zeno and the anti-Zeno regimes with the optimized measurements. 

%We additionally analyze another computational example(s) for different states. Consider the excited state of the two-level system, $\up$, specified by the angles $\theta = 0$ and $\phi = 0$. See FIG. \ref{SB3} for the graph. For this case, the graph for the pure dephasing model (orange large dashing) shows that the decay rate is zero. This make sense since the time evolution for the pure dephasing model changes only the diagonal terms of the density matrix. The population decay model (black dot-dashed) now exhibits both Zeno and anti-Zeno regimes -- since, after rotation about the $y$-axis, the state that is repeatedly measured (corresponding the original state) is a superposition of the $\up$ and $\down$ states. Since there is no (pure) dephasing, we now start observing lower decay rates at a smaller times if one makes optimal projective -- compared to other cases, as in FIG. \ref{SB1}. In this case, we observe differences in the decay rate starting from $\tau \approx 2$. The differences for times $\tau \approx 2$ are much larger as compared to the differences shown in the FIG. \ref{SB1}.

%\begin{figure}
%{\includegraphics[scale = 0.6]{SB3.eps}}\caption{\label{SB3} (Color Online) $\Gamma(\tau)$ %versus $\tau$ (low temperature) for the state specified %by $\theta = 0$ and $\phi = 0$. We use the same %parameters as in FIG. \ref{SB1}.}
%\end{figure}

\subsection*{Large Spin System}
\begin{figure}
{\includegraphics[scale = 0.55]{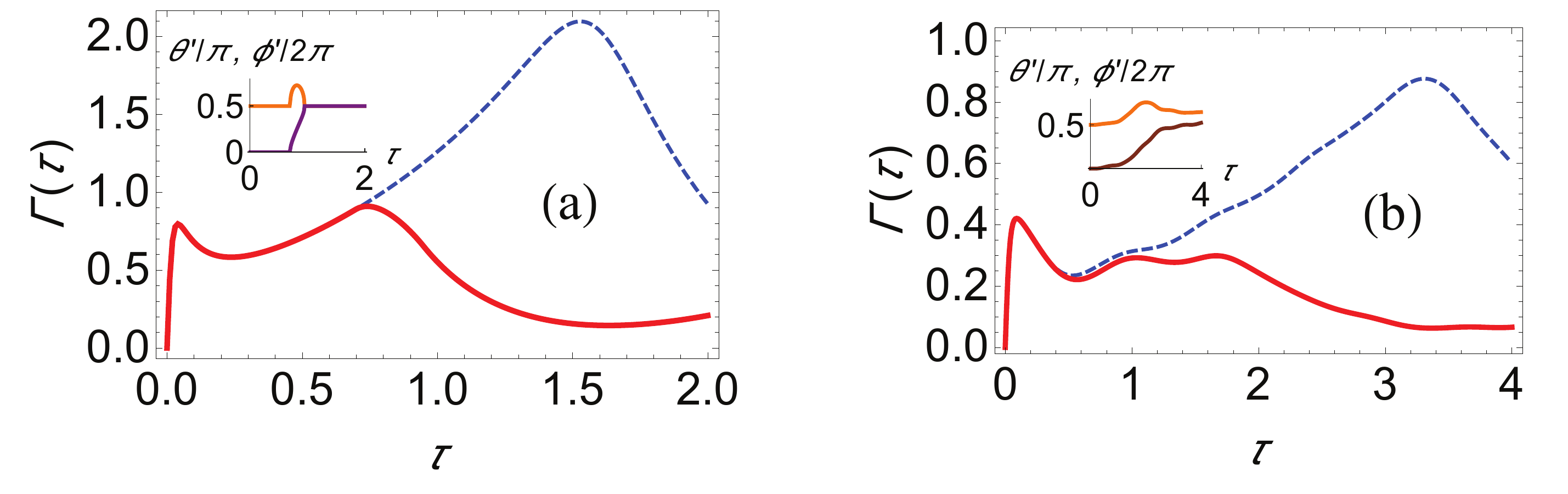}}
\caption{\label{LSJ1}\textbf{Graphs of the effective decay rate and the optimal spherical angles under making optimal projective measurements in the large spin model.} \textbf{(a)} $\Gamma(\tau)$ versus $\tau$ for $J=1$. Here we have $\Delta = 0$. The blue dashed curve shows the decay rate if the initial state is repeatedly measured; the red curve shows the decay rate if the optimal projective measurement is repeatedly made. We have used $G = 0.01, \omega_c = 50, \beta = 1$ and we take $\theta = \pi/2$ and $\phi = 0$ as parameters for the initial state. \textbf{(b)} Same as (a), except now that $\Delta \neq 0$. The insets show how the optimal measurements change with the measurement interval $\tau$.}
\end{figure}
We extend our study to a collection of two-level systems interacting with a common environment. This Hamiltonian can be considered to be a generalization of the usual spin-boson model to a large spin $j = N_s/2$ \cite{ChaudhryPRA2014zeno,VorrathPRL2005
,KurizkiPRL2011}, where $N_s$ is the number of two-level systems coupled to the environment. Physical realizations include a two-component Bose-Einstein condensate \cite{GrossNature2010,RiedelNature2010} that interacts with a thermal reservoir via collisions \cite{KurizkiPRL2011}. In this case, the system-environment Hamiltonian is given by 
\begin{equation}
H = \varepsilon J_{z} + \Delta J_x + \sum_{k} \omega_{k} b_{k}^{\dagger} b_{k} + 2 J_{z} \sum_{k} (g_{k}^{*}b_{k} + g_{k}b_{k}^{\dagger}),
\end{equation}
where $J_x$ and $J_z$ are the usual angular momentum operators and the environment is again modeled as a collection of harmonic oscillators. We first look at the pure dephasing case by settting $\Delta = 0$. In this case, the system dynamics can be found exactly. The system density matrix, in the eigenbasis of $J_z$, after removal of the evolution due to the system Hamiltonian can be written as 
$[\rho(t)]_{mn} = [\rho(0)]_{mn} e^{- i\triangle(t) (m^2 - n^2)} e^{- \gamma(t) (m - n)^2}$.
Here $\gamma(t)$ has been defined before, and $\Delta(t) = \sum_k 4|g_k|^2 \frac{[\sin(\omega_k t) - \omega_k t]}{\omega_k^2}$\text{\cite{ChaudhryPRA2014zeno}}
describes the indirect
interaction between the two-level systems due to their
interaction with a common environment. For
vanishingly small time $t$, $\triangle(t) \approx 0$. On the other
hand, as $t$ increases, the effect of 
$\triangle(t)$ becomes more
pronounced. Thus, we expect significant differences as compared to the single two-level system case for long measurement intervals. However, it is important to note that we can no longer find the optimal measurements using the formalism presented before since our system is no longer a single two-level system. In principle, we need then need to carry out a numerical optimization procedure in order to find the projector $\ket{\chi}\bra{\chi}$ such that the survival probability is maximized. Rather than looking at all possible states $\ket{\chi}$, we instead restrict ourselves to the SU(2) coherent states since these projective measurements are more readily experimentally accessible. In other words, we look at $\ket{\chi}\bra{\chi}$ where 
\begin{equation}
\ket{\chi} = \ket{\zeta, J} = (1 + |\zeta|^2)^{- J} \sum_{m=-J}^{m = J} \sqrt{\binom{2J}{J + m}} \zeta^{J + m} \ket{J, m},
\end{equation}
and $\zeta = e^{i \phi'} \tan(\theta'/2)$ with the states $\ket{J, m}$ being the angular momentum eigenstates of $J_z$. Suppose that we prepare the coherent state $\ket{\eta, J}$ with a fixed, pre-determined value of $ \eta = e^{i \phi} \tan(\theta/2)$ repeatedly. In order to do so, we project, with time interval $\tau$, the system state onto the coherent state $\ket{\zeta, J}$. After each measurement, we apply a suitable unitary operator to arrive back at the state $\ket{\eta,J}$. Again assuming the system-environment correlations are negligible, we find that 
\begin{align}\label{largedephasingdecay}
\Gamma(\tau) & = - \frac{1}{\tau} \ln \Bigg \{ \Bigg [ \frac{|\zeta|}{1 + |\zeta|^2} \Bigg ]^{2J} \Bigg [ \frac{|\eta|}{1 + |\eta|^2} \Bigg ]^{2J} \sum_{m, n = -J}^{J} (\zeta^{*} \eta)^m  (\eta^{*} \omega)^n \; \binom{2J}{J + m}\binom{2J}{J + n} e^{- i\triangle(\tau) (m^2 - n^2)} e^{- \gamma(\tau) (m - n)^2} \Bigg \}.
\end{align}
For equally spaced measurement time intervals, we numerically optimize Eq.~\eqref{largedephasingdecay} over the variables $\phi'$ and $\theta'$. We present a computational example in Fig.~\ref{LSJ1}(a). We take as the initial state the SU(2) coherent state with $\theta = \pi/2$ and $\phi = 0$ and we let $J = 1$. This is simply the generalization of the pure dephasing model that we have looked at before to $J = 1$. Previously, there was no difference in the optimized and unoptimized probabilities. Now, we see that because of the indirect interaction, there is a very noticeable difference. Where we observe the Zeno regime with the unoptimized measurements, we instead see the anti-Zeno regime with the optimized measurements. Furthermore, the survival probability can be significantly enhanced using the optimized measurements. 

For completeness, we have also considered the more general case with $\Delta \neq 0$. In this case, the system dynamics cannot be solved exactly, so we resort again to the master equation to find the system dynamics. With the system dynamics known, we again find the projector, parametrized by $\theta'$ and $\phi'$, such that the decay rate is minimized. Results are illustrated in ~\ref{LSJ1}(b). Once again, using optimizing projective measurements change the Zeno and anti-Zeno behaviour quatitatively as well as qualitatively.

\section*{Discussion}
Our central idea is that instead of repeatedly preparing the quantum state of a system using only projective measurements, we can repeatedly prepare the quantum state by using a combination of both projective measurements and unitary operations. This then allows us to consider the projective measurements that yield the largest survival probability. if the central quantum system is a simple two-level system, we have derived an expression that optimizes the survival probability, or equivalently the effective decay rate. This expression implies that the optimal projective measurement at time $\tau$ corresponds to the projector that is parallel to the Bloch vector of the system's density matrix at that time. We consequently applied our expression for the optimized survival probability to various models. For the population decay model, we found that beyond a critical time $\tau^{*}$, we should flip the measurement and start measuring the ground state rather than the excited state. For the pure dephasing model, we found that for states prepared in the equatorial plane of the Bloch sphere, it is optimal to measure the initial state - determining and making the optimal projective measurement has no effect on the effective decay rate. In contrast, for states prepared outside of the equatorial plane, the effect of making the optimal projective measurement substantially lowers the effective decay rate in the anti-Zeno regime. In the general spin-boson model, we have found that there can a considerable difference between the effective decay rate if we use the optimal measurements. We then extended our analysis by analyzing the case of large spin systems. In this case, we find that the indirect interaction between the two-level systems causes the optimal measurements to be even more advantageous. The results of this paper show that by exploiting the choice of the measurement performed, we could substantially decrease the effective decay rate for various cases. This allows us to `effectively freeze' the state of the quantum system with a higher probability. Experimental implementations
of the ideas presented in this paper are expected to be
important for measurement-based quantum control.

\end{document}